\documentclass[journal=jacsat,manuscript=article]{achemso}
\usepackage{upgreek}
\usepackage{dcolumn}
\usepackage{bm}
\usepackage{xcolor}
\usepackage{ulem}
\usepackage[T1]{fontenc}             
\usepackage[utf8]{inputenc}  
\usepackage[english]{babel}
\usepackage[babel]{csquotes}
\usepackage{newlfont}
\usepackage{color}

\usepackage{graphicx}

\usepackage{verbatim}
\usepackage{braket}
\usepackage[italian]{varioref}
\usepackage{float}
\usepackage{siunitx}
\usepackage{xcolor}
\usepackage[colorlinks=true,linkcolor=blue,citecolor=blue,urlcolor=blue]{hyperref}
\usepackage[all]{hypcap}
\usepackage{float}

\usepackage{upgreek}
\usepackage{dcolumn}
\usepackage{bm}
\usepackage{xcolor}
\usepackage{ulem}
\usepackage[T1]{fontenc}             
\usepackage[utf8]{inputenc}  
\usepackage[english]{babel}
\usepackage[babel]{csquotes}
\usepackage{newlfont}
\usepackage{color}

\usepackage{graphicx}

\usepackage{verbatim}
\usepackage{braket}
\usepackage[italian]{varioref}
\usepackage{float}
\usepackage{siunitx}
\usepackage{xcolor}
\usepackage[colorlinks=true,linkcolor=blue,citecolor=blue,urlcolor=blue]{hyperref}
\usepackage[all]{hypcap}
\usepackage{float}

\renewcommand{\suppinfo}{SM~\cite{supp-info}}
\newcommand{\qip}{\mathbf{q}}

\newcommand{\qc}{\mathbf{q}_\mathrm{c}}

\title{Excitonic effects in energy loss spectra of freestanding graphene}

\author{Alberto Guandalini}
\affiliation{S3 Centre, Istituto Nanoscienze, CNR, Via Campi 213/a, Modena (Italy)}
\alsoaffiliation{Dipartimento di Fisica, Universit\`a di Roma La Sapienza, Piazzale Aldo Moro 5, I-00185 Roma, Italy}
\email{alberto.guandalini@uniroma1.com}
\author{Ryosuke Senga}
\affiliation{Nanomaterials Research Institute, National Institute of Advanced Industrial Science and Technology (AIST), Tsukuba 305-8565, Japan}

\author{Yung-Chang Lin}
\affiliation{Nanomaterials Research Institute, National Institute of Advanced Industrial Science and Technology (AIST), Tsukuba 305-8565, Japan}

\author{Kazu Suenaga}
\affiliation{Nanomaterials Research Institute, National Institute of Advanced Industrial Science and Technology (AIST), Tsukuba 305-8565, Japan}
\alsoaffiliation{The Institute of Scientific and Industrial Research (SANKEN), Osaka University Mihogaoka 8-1, Ibaraki, Osaka 567-0047, Japan }

\author{Andrea Ferretti}
\affiliation{S3 Centre, Istituto Nanoscienze, CNR, Via Campi 213/a, Modena (Italy)}

\author{Daniele Varsano}
\affiliation{S3 Centre, Istituto Nanoscienze, CNR, Via Campi 213/a, Modena (Italy)}

\author{Andrea Recchia}
\affiliation{Center for Life NanoScience, Istituto Italiano di Tecnologia, viale Regina Elena 291, 00161 Rome, Italy}
\alsoaffiliation{Dipartimento di Fisica, Universit\`a di Roma La Sapienza, Piazzale Aldo Moro 5, I-00185 Roma, Italy}

\author{Paolo Barone}
\affiliation{CNR-SPIN, Area della Ricerca di Tor Vergata, Via del Fosso del Cavaliere 100, I-00133 Rome, Italy}
\alsoaffiliation{Dipartimento di Fisica, Universit\`a di Roma La Sapienza, Piazzale Aldo Moro 5, I-00185 Roma, Italy}

\author{Francesco Mauri}
\affiliation{Dipartimento di Fisica, Universit\`a di Roma La Sapienza, Piazzale Aldo Moro 5, I-00185 Roma, Italy}

\author{ Thomas Pichler}
\affiliation{University of Vienna, Faculty of Physics, Strudlhofgasse 4, A1090, Austria}

\author{Christian Kramberger}
\affiliation{University of Vienna, Faculty of Physics, Strudlhofgasse 4, A1090, Austria}

\begin{document}

\begin{abstract}
 In this work we perform electron energy-loss spectroscopy (EELS) of freestanding graphene with high energy and momentum resolution to disentangle the quasielastic scattering from the excitation gap of Dirac electrons close to the optical limit. We show the importance of many-body effects on electronic excitations at finite transferred momentum by comparing measured EELS with ab initio calculations at increasing levels of theory. Quasi-particle corrections and excitonic effects are addressed within the GW approximation and Bethe-Salpeter equation, respectively.
Both effects are essential in the description of the EEL spectra to obtain a quantitative agreement with experiments, with the position, dispersion, and shape of both the excitation gap and the $\pi$ plasmon being significantly affected by excitonic effects.

\end{abstract}

\maketitle
\section{Introduction}

Since its discovery~\citep{Novoselov_2016,novoselov_two-dimensional_2005}, the electronic excitations in graphene have been extensively studied because of their relevance for  plasmonics  and optoelectronics~\citep{CastroNeto_rev2009,Bonaccorso_2010,Grigorenko_2012, Garcia_2014}. Electron energy loss  (EEL) spectroscopy (EELS) resolved both in energy and momentum is a powerful tool to investigate dispersion relations of electronic excitations, as it directly probes the dynamical inverse dielectric function (i.e.\ the energy-loss function) which provides information on both collective longitudinal excitations (plasmons) and electron-hole pair excitations. The energy loss function of undoped graphene and of related $sp^2$ carbon materials, graphite and nanotubes, always display $\pi$ and $\pi+\sigma$ plasmons, arising from interband transitions occurring below 10~eV and above 15~eV, respectively~\citep{Taft1965,Zeppenfeld1967,Kinyanjui2012,Wachsmuth2013,Wachsmuth2014,Kramberger2008}. Dimensionality affects their dispersion relations via interlayer Coulomb interaction, resulting in a parabolic dependence on the momentum of the $\pi$ plasmon in  graphite~\citep{Zeppenfeld1967}, as opposed to its linear dispersion measured in single-wall nanotubes~\citep{Kramberger2008} and freestanding single-layer graphene~\cite{Kinyanjui2012,Wachsmuth2013}. 

Besides plasmons, undoped graphene displays electron-hole pair excitations due to interband transitions between lower and upper Dirac cones~\citep{CastroNeto_rev2009}, resulting in a momentum-dependent onset of the EEL spectra. Its direct measurement has been so far elusive due to the background of non-scattered electrons, concealing low-energy features within the zero-loss peak in experiments. On the other hand, interband transitions have been experimentally inferred from the linewidth of graphite surface plasmons, decreasing with increasing in-plane momentum transfer as the electron-hole continuum is more and more gapped, thus reducing the number of possible plasmon decay-channels into electron-hole  excitations~\cite{Laitenberger1996}.

Inelastic X-ray scattering measurements of graphite have also been used
to extract longitudinal excitations in graphene by
removing the Coulomb interaction between the layers from the graphite
response~\citep{Reed_2010,Gan_2016}. However, low-energy excitonic effects in graphite are
expected to be smaller than in graphene, as the interaction between
electrons and holes is more efficiently
screened by the multiple layers~\citep{Trevisanutto_2010_b}. Thus, a proper estimation of
excitonic effects in graphene
can be made only by directly measuring (or calculating) the response
properties of the 2D layer surrounded by vacuum.

From a theoretical perspective, the dielectric response of freestanding graphene has been extensively studied with a variety of first-principles approaches based on density functional theory (DFT), and including many-body effects at different levels of theory. Starting from a DFT band structure, the energy-loss function at finite momentum transfer has been calculated mostly within the random-phase approximation (RPA)~\citep{Wachsmuth2013,Wachsmuth2014,Kramberger2008,Yan_prl2011,Despoja_2012,Despoja_2013,Mowbray_2014,Novko_2015,Nazarov_2015,Li_2017}
{\color{orange}\citep{Zheng_2017}}.
DFT+RPA calculations accurately reproduce the experimental linear dispersion of the $\pi$ plasmon only for momentum transfers along the graphene plane $ \qip > \qc \simeq 0.1$ \AA$^{-1}$, both in the $\Gamma K$ and the $\Gamma M$ directions~\cite{Wachsmuth2013,Wachsmuth2014}.

Many-body effects stemming from electron-electron (e-e) and electron-hole (e-h) interactions have been analyzed, only for optical spectra, namely for vanishing transferred momentum~\citep{Trevisanutto_2008,Yang2009,Yang2011}, within the GW approximation~\citep{Hedin_1965,Strinati_1982,Hybertsen_1986,Godby_1988}, accounting for repulsive e-e coupling in the band structure, and solving the Bethe-Salpeter equation (BSE)~\citep{Hedin_1965,Strinati_1988,Onida_2002} to include attractive e-h interactions.  
The inclusion of e-e effects strongly modifies the quasi-particle (QP) band structure, affecting not only the Dirac Fermi-velocity but also the energy gaps at the $M$ and $\Gamma$ points, and consequently the spectral position of the onset as well as of the $\pi$ and $\pi+\sigma$ plasmons. Indeed, inclusion of QP corrections within the GW approach has proven crucial to predict Fermi velocities in better agreement with ARPES~\citep{Trevisanutto_2008,Yang2009,Attaccalite_2009}. On the other hand,
the larger QP gap at the $M$ point due to e-e effects alone is responsible for a substantial blueshift of the $\pi$-plasmon peak in the calculated optical spectrum of graphene, with the predicted spectral position exceeding  the experimental one by
roughly 600~meV~\citep{Trevisanutto_2008,Yang2009,Yang2011}. As shown by Yang et al.~\citep{Yang2009,Yang2011}, for zero-momentum transfer the explicit inclusion of e-h coupling via BSE is then essential for reproducing the experimental peak position as well as the asymmetric profile of the excitations that gives rise to the $\pi$ plasmon in EELS.

The importance of many-body and excitonic effects in the dielectric response of freestanding graphene can be anticipated by considering the reduced screening due to both its vanishing density of states at the Fermi level and its two-dimensional character~\citep{Thygesen_2017}. While the good agreement of DFT+RPA calculations of the $\pi$ plasmon dispersion with experiments for $ \qip > \qc$ has been tentatively ascribed to a cancellation of e-e and e-h effects~\citep{Wachsmuth2013}, it is still unknown how the sizeable excitonic and many-body effects that have been unveiled in the optical spectrum evolve in the finite-momentum regime. 

In this work, we precisely address the regime of small, but finite, momentum transfers.
EELS measurements of freestanding graphene have been performed in a transmission electron microscope (TEM) with unprecedented resolution both in energy and momentum to clearly separate 
the zero loss peak (ZLP) from the excitation gap of Dirac electrons.
We focused on the low-energy (0-10 eV) EEL spectra in the $\qip\le0.3$~\AA$^{-1}$ range of momentum transfers. With the adopted setup, we were able to access not only the collective $\pi$-plasmon excitation, but also the onset of the particle-hole continuum, thus providing complementary information about excitonic effects on the two main spectral features in the considered momentum-energy range. 
Momentum-resolved EEL spectra have been simulated with first principles methods at increasing levels of theory, from the single-particle picture of the DFT+RPA calculations to GW computations accounting for many-body QP corrections, finally addressing excitonic effects and e-h interactions within BSE. This allowed us to assess the effects of e-e and e-h interactions separately by comparing spectra simulated at different approximation levels. Our comparative analysis unveils non-trivial momentum-dependent excitonic effects in the dielectric response of graphene, whose signatures shape differently the EEL spectral features of both electron-hole pair and collective excitations.

\begin{figure}
\includegraphics[width=\linewidth]{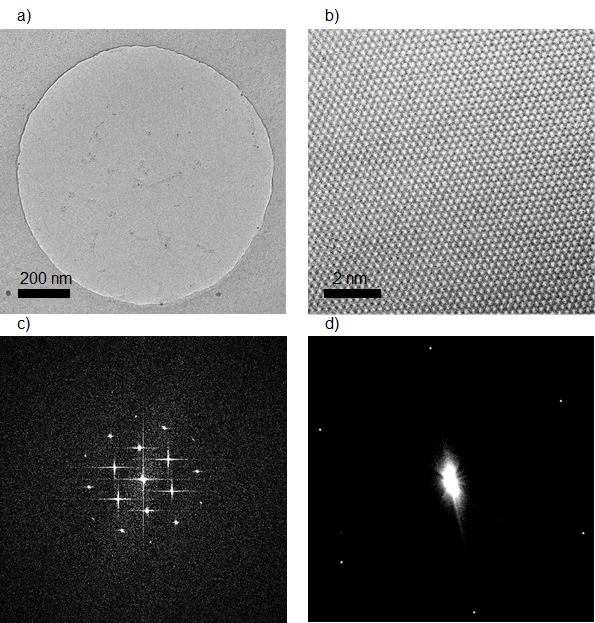}
\caption{ TEM images and diffraction pattern of freestanding graphene at $30$~keV. (a): low-magnification TEM image. (b): high resolution TEM image. (c): FFT of the high resolution image in (b). (d): diffraction pattern of a $500$~nm wide beam acquired at a camera length of $40$~cm.}
\label{Fig_TEM}
\end{figure}

\begin{figure}
\includegraphics[width=\linewidth]{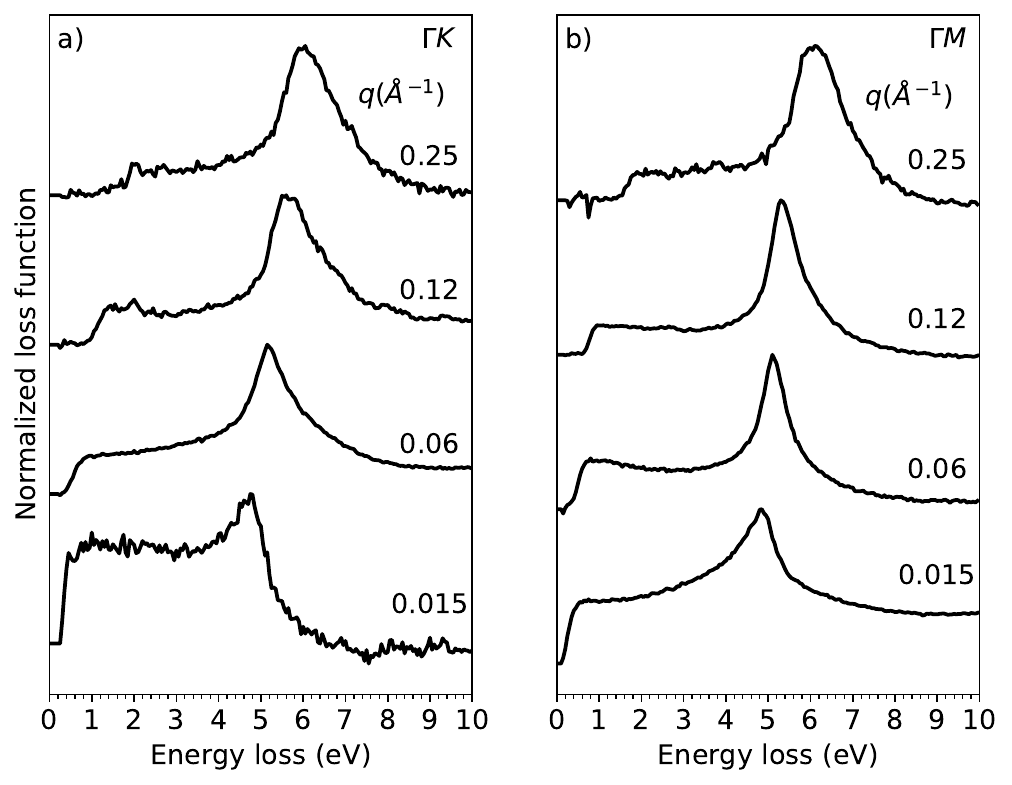}
\caption{Measured low energy loss spectra of freestanding graphene for different momentum transfers $\qip$ oriented along the $\Gamma K$(a) and $\Gamma M$(b) directions. The zero loss peak has been removed (see text).
 The complete set of measured spectra is included in the \suppinfo.
}
\label{Fig_exp}
\end{figure}
%

\section{Methods}
Single layer graphene was synthesized by plasma-assisted chemical vapour deposition \cite{Kato2016}. The graphene layers were mechanically transferred to TEM grids. The samples were annealed \textit{in situ} at 500$^\circ$C overnight and  kept at the same temperature during all measurements  to mantain crystallinity and avoid surface contamination.
 The low magnification TEM image of the graphene on the TEM grid exhibits a free-standing area with a hole diameter of approximately 1 $\mu$m [Fig.~\ref{Fig_TEM}(a)]. The typical crystal structure of the sample and its Fast Fourier Transform (FFT) image are presented in Fig.~\ref{Fig_TEM} (b) and (c), respectively. Both images show a defect free single crystal in a $10\times10$~nm$^2$ wide region. Fig.~\ref{Fig_TEM} (d) is the lower magnified overview diffraction image that was used to position the aperture in the diffraction plane. The positioning and beam drift control before and after each acquisition are checked in the low magnification diffraction pattern. They are intrinsically limited to the pixel size of 0.004~\AA$^{-1}$. The spectral resolution of $\pm0.02~\text{\AA}^{-1}$ is achieved with a longer camera length (215~cm).
EEL spectra were measured at 30~kV in a JEOL TEM (3C2) equipped with a Schottky field emission gun, a double Wien filter monochromator and delta correctors. The energy resolution was set to $45$~meV in full width half maximum (FWHM). A collimated 500~nm wide beam ($\qip\leq 0.006$~\AA$^{-1}$) was formed in imaging mode. The incident beam was perpendicular to the graphene layer, and the diffraction plane was formed with a camera length of 215~cm. A pin-hole type aperture ($\pm$0.015~\AA$^{-1}$) was inserted in the diffraction plane.
The beam current was approximately 10~pA. The EEL spectra were collected from multiple $\qip$ along ${\Gamma K}$ and ${\Gamma M}$ with a low-voltage optimized GATAN GIF quantum spectrometer. 
%
The spectra are sums of 300 individual acquisitions. The dwell time ranges between 0.01~s (for $\qip$=0.015~\AA$^{-1}$) to 2 sec (for $\qip$=0.3~\AA$^{-1}$).
The zero-loss peak (ZLP) is clearly separated from the onset (i.e.\ the excitation gap) for $\qip>$0.05~\AA$^{-1}$. At smaller $\qip$, the onset overlap with the tails of the ZLP.
At all momentum transfers, we modelled the ZLP with an inverse power law\cite{Krivanek_2014,Hatchel_2018}, and removed as described in the supplementary material (SM)~\cite{supp-info}. In Fig.~\ref{Fig_gap}, black stars correspond to data where the ZLP is clearly separated from the onset, while grey stars ($\qip<$0.05~\AA$^{-1}$) correspond to onets that have only become visible after the ZLP subtraction, and their uncertainties are on the order of the onset itself.

DFT calculations were performed using a plane wave basis set as implemented in the {\sc Quantum ESPRESSO} package~\citep{QE_2020}, with the local density approximation~\citep{LDA}.
The {\sc Yambo} code~\citep{yambo_2009,yambo_2019} has been adopted
to compute the quasi-particle band structure, within the $G_0W_0$ approximation, and the EEL spectra, calculated at different levels of theory, considering DFT and GW QP energy levels within the RPA and the BSE response functions. In all cases, finite-$\qip$ simulations have been performed.
The adopted supercell includes a vacuum region along the perpendicular direction of 10~\AA, and a 2D slab Coulomb  cutoff~\citep{Beigi_2006,Rozzi_2006} has been used to avoid spurious interactions. 
For GW calculations, we adopted
the plasmon-pole approximation in the Godby-Needs scheme~\citep{Godby_1989}.
The long-wavelength limit of the semimetallic screening contribution have been included analytically within the Dirac cone model (see \suppinfo).
Excitation broadening, mainly originating from electron-phonon coupling, has been included through the model described in Ref.~\citep{Venezuela2011}.
Further details can be found in the~\suppinfo.

\section{Results}
The measured low-energy (0-10~eV) loss spectra are shown in Fig.~\ref{Fig_exp} for selected momentum transfers $\qip$ along the $\Gamma K$ and $\Gamma M$ directions.
As expected, a peak due to the $\pi$ plasmon is present at about $5$-$6$ eV. Notably, a clear finite momentum-dependent onset is present in all the measured spectra. To our knowledge, this is the first direct experimental evidence of such onset dispersion for freestanding graphene at such low momentum transfers (see below for a detailed discussion). 
Interestingly, the line shape of the $\pi$ plasmon shows a transition from an asymmetric profile with a tail towards lower energies in the low momentum-transfer regime, to a symmetric shape around $\qip$ = 0.06~\AA$^{-1}$, and finally a reversed asymmetric profile leaning towards higher energies 
for $\qip > \qc$.
Spectral features are found to be essentially independent of the crystallographic direction in the measured momentum range, consistent with previous reports~\cite{Kinyanjui2012,Wachsmuth2014}.

\begin{figure}
\includegraphics[width= \linewidth]{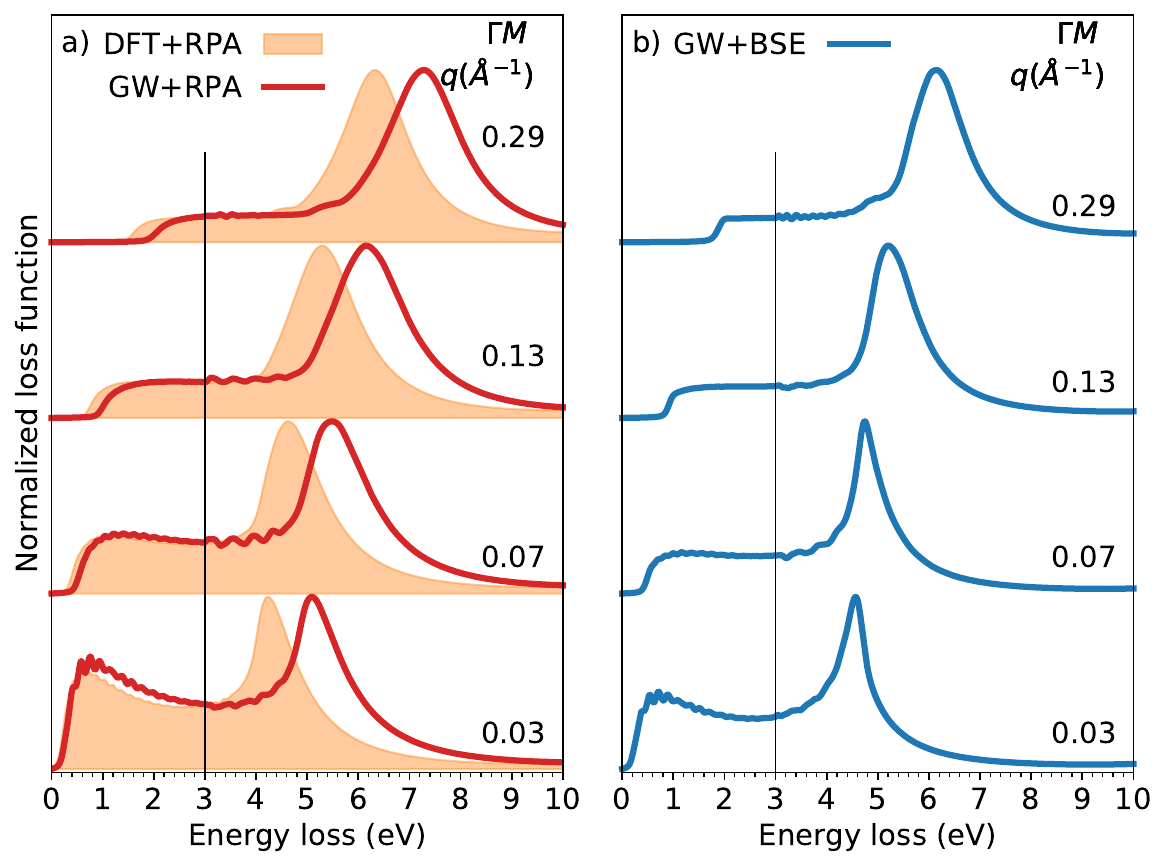}
\caption{Energy loss spectra of freestanding graphene at increasing momentum transfers along 
${\Gamma M}$
calculated at different theory levels: DFT+RPA and GW+RPA (a), GW+BSE (b).
For all the spectra, an energy dependent broadening has been considered to describe relaxation processes due to electron-phonon coupling, according to Eqn.~(14) of Ref.~\citep{Venezuela2011}. 
The vertical black lines in the panels (b,c) separate the regions for different convergence parameters (see text). }
\label{Fig_calc}
\end{figure}

In order to investigate the origin of the observed experimental features, we compare the measured data with first principle calculations of the loss function.
In Fig.~\ref{Fig_calc}, we show EEL spectra calculated at different levels of theory along the $\Gamma M$ direction  (calculations for more $\qip$ are included in the \suppinfo).
The first$+$second acronym in the labels refers to the level of theory used for the band structure$+$response function.
In panel (a) we report the loss function computed at the DFT+RPA level (orange shaded area), which, as an effective single particle picture, we then use as a starting point to understand many-body effects.
The inclusion of GW corrections in the band structure (GW+RPA), red line in panel (a), result in an almost rigid blue-shift of the $\pi$-plasmon peak. Instead, when including e-h coupling effects using GW+BSE, the peaks are  shifted and the shape is modified in a non uniform way, as clearly seen in Fig.~\ref{Fig_calc}(b). 
In particular, one can
recognize that the experimentally observed changes in the asymmetry of the plasmon peak can only be reproduced by including electron-hole interaction effects (GW+BSE).

\begin{figure}
\includegraphics[width=\linewidth]{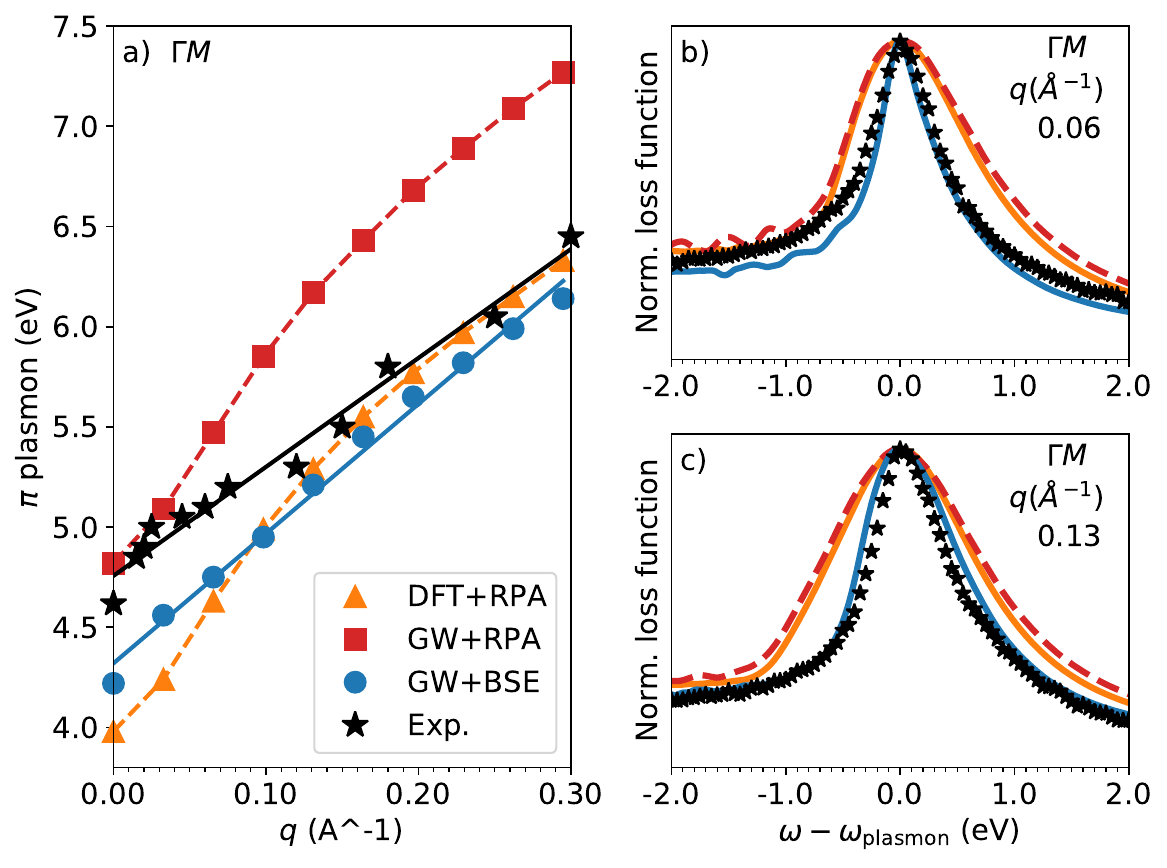}
\caption{
(a): $\pi$-plasmon dispersion obtained with the procedure described in the \suppinfo.
The straight line in black (blue) is a linear regression of the measured (calculated) plasmon dispersion. Dashed lines are a guide to the eye.
(b, c): measured and calculated $\pi$-plasmon peaks of freestanding graphene 
at different momentum transfers along the $\Gamma M$ direction.
(b): $\qip = 0.06$ \AA$^{-1}$; (c): $\qip = 0.13$ \AA$^{-1}$.
Different colours refer to different theory levels in the calculation: DFT+RPA (orange lines), GW+RPA (red lines) and GW+BSE (blue lines).  Experimental data are shown as black stars.
\label{Fig_pl}}
\end{figure}

This scenario is even more evident by looking at Fig.~\ref{Fig_pl}, where we compare the features of the $\pi$ plasmon obtained experimentally and from \textit{ab initio} calculations.
The independent particle picture (DFT+RPA) is not able to reproduce the experimental dispersion below $\qc\simeq0.1$~\AA$^{-1}$, as already pointed out in Refs.~\cite{Wachsmuth2013, Wachsmuth2014}.
The DFT+RPA dispersion is in fact quadratic in the low-momentum regime, transitioning to a linear behaviour closely matching to experiments for $\qip > \qc$.
The GW corrections to the band structure near $M$ (i.e.\ the relevant BZ region for what concerns the $\pi$ plasmon) enlarge the gap by about 0.88~eV.
Thus, the inclusion of e-e interaction within GW+RPA determines a rigid blue shift that reproduces well the experimental optical limit, but fails for finite $\qip$ and in keeping a parabolic dispersion for small momenta.

Excitonic effects (included via BSE) 
produce instead a momentum-dependent red shift, that 
increases with $\qip$ before it saturates around $\qc$.
Such a momentum-dependent effect due to e-h interaction is responsible for the linear dispersion of the $\pi$ plasmon in the GW+BSE results (blue symbols and line in Fig.~\ref{Fig_pl}a), which reproduces the experimental trend well.
In Fig.~\ref{Fig_pl}(b,c) we compare the measured and calculated shapes of the $\pi$-plasmon peak at $\qip = 0.06$ \AA$^{-1}$ and $\qip = 0.13$ \AA$^{-1}$, respectively, by shifting the plasmon peak so that its center is located at $\omega = 0$.
DFT-RPA data show in both cases a peak that is too broad, with less asymmetric features. The inclusion of the e-e interaction (GW+RPA) does not influence the peak shape.
Instead, the peak reshaping is a feature provided by the e-h interaction, as it can be seen from the nearly perfect match between experimental and GW+BSE results.

\begin{figure}
\includegraphics[width=\linewidth]{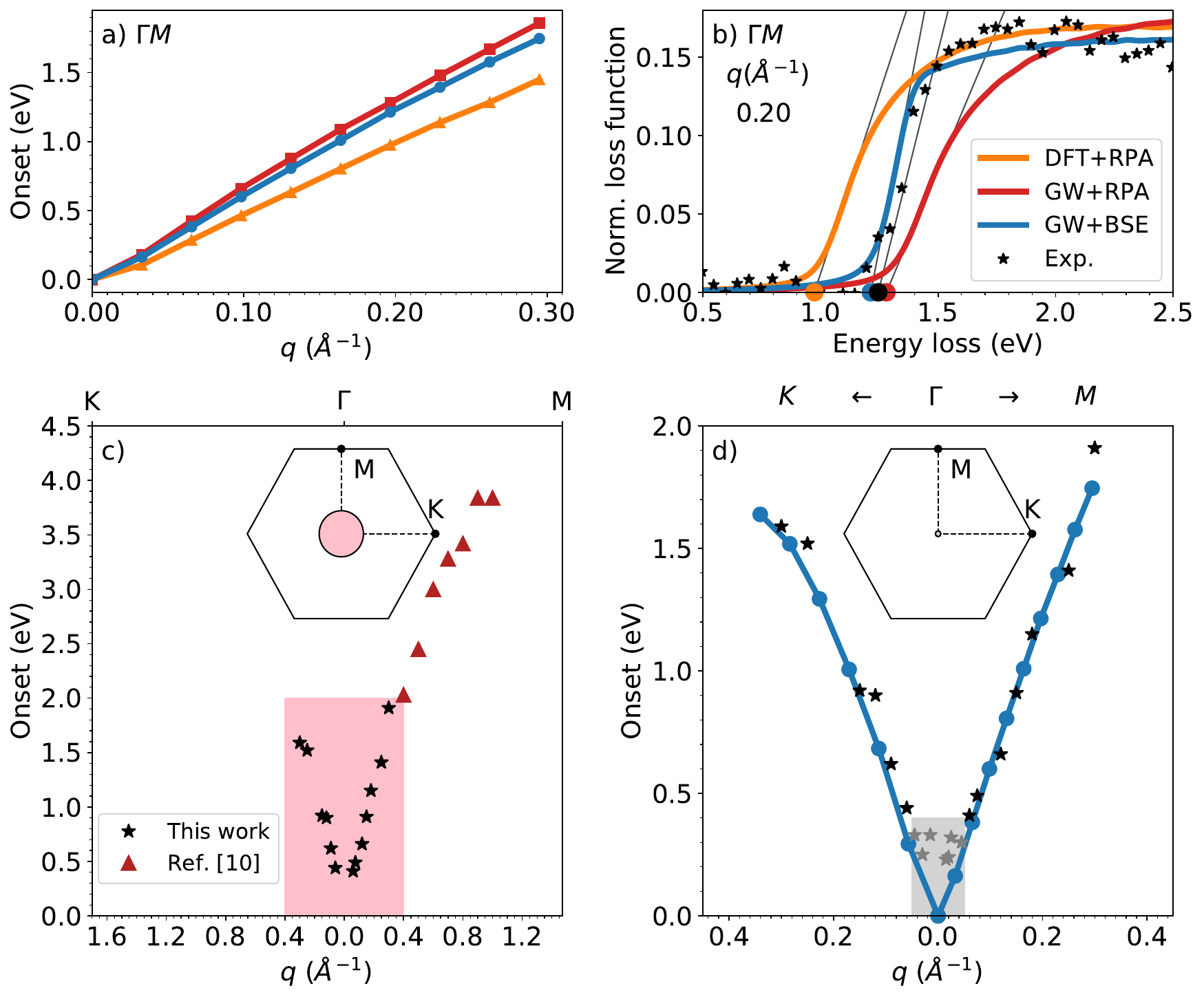}
\caption{
(a): onset dispersion of freestanding graphene  along the $\Gamma M$ direction calculated at different theory levels. (b): 
experimental and calculated onset shape at $\qip \approx 0.2$ \AA$^{-1}$.
Grey lines indicate the half-height tangent used in the determination of the onset. Filled circles indicate the onset positions. 
(c): measured onsets for $\qip$ in $\Gamma K$ and $\Gamma M$ directions.
We extracted from Ref.\citep{Wachsmuth2013} onset positions at high momentum transfer along the $\Gamma M$ direction.
The extraction has not been done in the $\Gamma K$ direction due to experiental uncertainties.
The onset inaccessible region of Ref.\citep{Wachsmuth2013} ($\qip\le0.4$)~\AA$^{-1}$ has been tinted in red. It is also shown inside the first BZ of graphene.
(d): zoom of the measured onset dispersion (black) and comparison with the calculated 
onset (blue). Our range of inaccessible $\qip\le0.05$)~\AA$^{-1}$ is tinted in grey and is also shown within the first BZ.
The same region has been sketched on the first BZ of graphene. The grey data points within $\qip\le0.05$~\AA$^{-1}$ have large errors as they depend on the removal of the ZLP.
}
\label{Fig_gap} 
\end{figure}
%

We are now in the position to discuss the low energy features of the EELS data, notably including the onset dispersion observed experimentally (see Fig.~\ref{Fig_exp}).
In particular, in Fig.~\ref{Fig_gap} we analyse the position and shape of the onset spectral structures, related to the e-h pair excitation gap obtained at finite momentum transfer $\qip$, as expected from the semimetallic nature of freestanding graphene. 
The position of the onset is extracted from experimental data by considering the tangent at half-hight of the onset structure, and then taking the intercept with the horizontal axis (see Fig.~\ref{Fig_gap}b). The same procedure is applied also to the theoretical data.
With these definitions, it is evident (Fig.~\ref{Fig_gap}a,b) that the onset position is sensibly dependent on the choice of the theory level used to describe the band structure. 
Indeed, the inclusion of e-e interactions (GW corrections) significantly enlarge the gap opening with a higher Fermi velocity.

In Fig.~\ref{Fig_gap}(b) we show the onset shape at $\qip = 0.20$ \AA$^{-1}$. While the onset position is similar in the GW+RPA and GW+BSE cases, it is evident that e-h interaction modifies the onset shape, providing a better agreement with experiments. 
We have explicitly verified that this feature mainly originates from matrix element effects, i.e.\ the excitonic eigenvalues (DOS) are mostly unchanged while the red shift comes from a redistribution of their spectral weights.
Also in this case, the GW+BSE calculations consistently produce  dispersions in excellent agreement with experiments, as shown in Fig.~\ref{Fig_gap}(d). In Fig.~\ref{Fig_gap}(c) we compare our current onset positions with those extracted from Ref.\citep{Wachsmuth2013}. The red disk in the sketched Brillouin zone (BZ) marks the range $\qip<0.4$~\AA$^{-1}$, where the spectral onset could not be accessed in Ref.\citep{Wachsmuth2013}. Here we pushed this accessibility limit down to $\qip\le0.05$~\AA$^{-1}$ (grey disk and range) in Fig.~\ref{Fig_gap}(d).
The measured onsets with $\qip\le0.05$~\AA$^{-1}$ are dependent on the details of the ZLP subtraction and marked grey. Their errors are comparable to the gap opening.
The anisotropy in the onset dispersion along the $\Gamma M$ and $\Gamma K$ directions in panels (c) and (d) arises from the trigonal warping of the Dirac cones. 

\section{Conclusions}   
 We performed momentum-resolved EELS measurements of freestanding graphene in a TEM with a high energy and momentum resolution, that is the key to disentangle the near optical excitations of Dirac electrons from the quasielastic scattering in the low momentum-transfer regime, allowing us to resolve a number of spectral features as well as the dispersive nature of excitonic effects in graphene.
Jointly, we perform \textit{ab initio} calculations 
of the loss function at different levels of theory, ranging from DFT+RPA to GW+BSE.
The comparison between measurements and \textit{ab initio} calculations allows us to unravel the physical origins behind the observed spectral features, namely position and shape of the $\pi$-plasmon peak and the spectral onset.  

For what concerns the $\pi$ plasmon dispersion, we found that the inclusion of electron-electron interactions at the GW level provides an almost rigid blue-shift of the peaks that is independent of the momentum transfer, such that this correction alone does not help in
reproducing the experimentally observed features.
Instead, the further inclusion of electron hole-attraction provides an unexpected momentum-dependent red-shift, that is responsible for the 
linear plasmon peak dispersion clearly observed in experiments.
The asymmetric shapes of the $\pi$-plasmon peak can also only be ascribed to excitonic effects, as already shown in optics.
The dispersion of the onset as a function of the transferred momentum is already in quantitative agreement with experiments once the electron-electron interaction is included. 
However, excitonic effects still play a fundamental role in describing the onset shape.

\begin{acknowledgement}
RS, YCL and KS acknowledge the support from JST-CREST (JPMJCR20B1, JPMJCR20B5, JPMJCR1993), JST-PRESTO (JPMJPR2009), and JSPS-KAKENHI (JP21H05235, JP22H05478, JP23H00277). This project has received funding from the European Research Council (ERC) under the European Union's Horizon 2020 research and innovation program (MORE-TEM ERC-SYN project, grant agreement No 951215).
DV, AF and AG acknowledge the support from MaX -- MAterials design at the eXascale -- a European Centre of Excellence funded by the European Union's program 
HORIZON-EUROHPC-JU-2021-COE-01 (Grant No. 101093374), 
ICSC – Centro Nazionale di Ricerca in High Performance Computing, Big Data and Quantum Computing, funded by European Union –NextGenerationEU - PNRR, Missione 4 Componente 2 Investimento 1.4 and the Italian national program PRIN2017 2017BZPKSZ “Excitonic insulator in two-dimensional long-range interacting systems".
We acknowledge CINECA for computational resources, awarded via the ISCRA Grants No. HP10BKBJMI and HP10BV0TBS

\end{acknowledgement}

\bibliography{UltraResolutionGap}

\providecommand{\latin}[1]{#1}
\makeatletter
\providecommand{\doi}
  {\begingroup\let\do\@makeother\dospecials
  \catcode`\{=1 \catcode`\}=2 \doi@aux}
\providecommand{\doi@aux}[1]{\endgroup\texttt{#1}}
\makeatother
\providecommand*\mcitethebibliography{\thebibliography}
\csname @ifundefined\endcsname{endmcitethebibliography}
  {\let\endmcitethebibliography\endthebibliography}{}
\begin{mcitethebibliography}{48}
\providecommand*\natexlab[1]{#1}
\providecommand*\mciteSetBstSublistMode[1]{}
\providecommand*\mciteSetBstMaxWidthForm[2]{}
\providecommand*\mciteBstWouldAddEndPuncttrue
  {\def\EndOfBibitem{\unskip.}}
\providecommand*\mciteBstWouldAddEndPunctfalse
  {\let\EndOfBibitem\relax}
\providecommand*\mciteSetBstMidEndSepPunct[3]{}
\providecommand*\mciteSetBstSublistLabelBeginEnd[3]{}
\providecommand*\EndOfBibitem{}
\mciteSetBstSublistMode{f}
\mciteSetBstMaxWidthForm{subitem}{(\alph{mcitesubitemcount})}
\mciteSetBstSublistLabelBeginEnd
  {\mcitemaxwidthsubitemform\space}
  {\relax}
  {\relax}

\bibitem[Novoselov \latin{et~al.}(2004)Novoselov, Geim, Morozov, Jiang, Zhang,
  Dubonos, Grigorieva, and Firsov]{Novoselov_2016}
Novoselov,~K.~S.; Geim,~A.~K.; Morozov,~S.~V.; Jiang,~D.; Zhang,~Y.;
  Dubonos,~S.~V.; Grigorieva,~I.~V.; Firsov,~A.~A. Electric Field Effect in
  Atomically Thin Carbon Films. \emph{Science} \textbf{2004}, \emph{306},
  666--669\relax
\mciteBstWouldAddEndPuncttrue
\mciteSetBstMidEndSepPunct{\mcitedefaultmidpunct}
{\mcitedefaultendpunct}{\mcitedefaultseppunct}\relax
\EndOfBibitem
\bibitem[Novoselov \latin{et~al.}(2005)Novoselov, Geim, Morozov, Jiang,
  Katsnelson, Grigorieva, Dubonos, and Firsov]{novoselov_two-dimensional_2005}
Novoselov,~K.~S.; Geim,~A.~K.; Morozov,~S.~V.; Jiang,~D.; Katsnelson,~M.~I.;
  Grigorieva,~I.~V.; Dubonos,~S.~V.; Firsov,~A.~A. Two-dimensional gas of
  massless {Dirac} fermions in graphene. \emph{Nature} \textbf{2005},
  \emph{438}, 197--200\relax
\mciteBstWouldAddEndPuncttrue
\mciteSetBstMidEndSepPunct{\mcitedefaultmidpunct}
{\mcitedefaultendpunct}{\mcitedefaultseppunct}\relax
\EndOfBibitem
\bibitem[Castro~Neto \latin{et~al.}(2009)Castro~Neto, Guinea, Peres, Novoselov,
  and Geim]{CastroNeto_rev2009}
Castro~Neto,~A.~H.; Guinea,~F.; Peres,~N. M.~R.; Novoselov,~K.~S.; Geim,~A.~K.
  The electronic properties of graphene. \emph{Rev. Mod. Phys.} \textbf{2009},
  \emph{81}, 109--162\relax
\mciteBstWouldAddEndPuncttrue
\mciteSetBstMidEndSepPunct{\mcitedefaultmidpunct}
{\mcitedefaultendpunct}{\mcitedefaultseppunct}\relax
\EndOfBibitem
\bibitem[Bonaccorso \latin{et~al.}(2010)Bonaccorso, Sun, Hasan, and
  Ferrari]{Bonaccorso_2010}
Bonaccorso,~F.; Sun,~Z.; Hasan,~T.; Ferrari,~A.~C. Graphene photonics and
  optoelectronics. \emph{Nature Photonics} \textbf{2010}, \emph{4}, 611\relax
\mciteBstWouldAddEndPuncttrue
\mciteSetBstMidEndSepPunct{\mcitedefaultmidpunct}
{\mcitedefaultendpunct}{\mcitedefaultseppunct}\relax
\EndOfBibitem
\bibitem[Grigorenko \latin{et~al.}(2012)Grigorenko, Polini, and
  Novoselov]{Grigorenko_2012}
Grigorenko,~A.; Polini,~M.; Novoselov,~K. Graphene plasmonics. \emph{Nature
  Photon.} \textbf{2012}, \emph{6}, 749--758\relax
\mciteBstWouldAddEndPuncttrue
\mciteSetBstMidEndSepPunct{\mcitedefaultmidpunct}
{\mcitedefaultendpunct}{\mcitedefaultseppunct}\relax
\EndOfBibitem
\bibitem[García~de Abajo(2014)]{Garcia_2014}
García~de Abajo,~F.~J. Graphene Plasmonics: Challenges and Opportunities.
  \emph{ACS Photonics} \textbf{2014}, \emph{1}, 135--152\relax
\mciteBstWouldAddEndPuncttrue
\mciteSetBstMidEndSepPunct{\mcitedefaultmidpunct}
{\mcitedefaultendpunct}{\mcitedefaultseppunct}\relax
\EndOfBibitem
\bibitem[Taft and Philipp(1965)Taft, and Philipp]{Taft1965}
Taft,~E.~A.; Philipp,~H.~R. {Optical Properties of Graphite}. \emph{Phys. Rev.}
  \textbf{1965}, \emph{138}, A197--\&\relax
\mciteBstWouldAddEndPuncttrue
\mciteSetBstMidEndSepPunct{\mcitedefaultmidpunct}
{\mcitedefaultendpunct}{\mcitedefaultseppunct}\relax
\EndOfBibitem
\bibitem[Zeppenfeld(1967)]{Zeppenfeld1967}
Zeppenfeld,~K. {Anisotropic Plasmon Behaviour in Graphite}. \emph{Phys. Lett.
  A} \textbf{1967}, \emph{25}, 335--+\relax
\mciteBstWouldAddEndPuncttrue
\mciteSetBstMidEndSepPunct{\mcitedefaultmidpunct}
{\mcitedefaultendpunct}{\mcitedefaultseppunct}\relax
\EndOfBibitem
\bibitem[Kinyanjui \latin{et~al.}(2012)Kinyanjui, Kramberger, Pichler, Meyer,
  Wachsmuth, Benner, and Kaiser]{Kinyanjui2012}
Kinyanjui,~M.~K.; Kramberger,~C.; Pichler,~T.; Meyer,~J.~C.; Wachsmuth,~P.;
  Benner,~G.; Kaiser,~U. Direct probe of linearly dispersing 2D interband
  plasmons in a free-standing graphene monolayer. \emph{Europhys. Lett.}
  \textbf{2012}, \emph{97}, 57005\relax
\mciteBstWouldAddEndPuncttrue
\mciteSetBstMidEndSepPunct{\mcitedefaultmidpunct}
{\mcitedefaultendpunct}{\mcitedefaultseppunct}\relax
\EndOfBibitem
\bibitem[Wachsmuth \latin{et~al.}(2013)Wachsmuth, Hambach, Kinyanjui, Guzzo,
  Benner, and Kaiser]{Wachsmuth2013}
Wachsmuth,~P.; Hambach,~R.; Kinyanjui,~M.~K.; Guzzo,~M.; Benner,~G.; Kaiser,~U.
  {High-energy collective electronic excitations in free-standing single-layer
  graphene}. \emph{Phys. Rev. B} \textbf{2013}, \emph{88}, 075433\relax
\mciteBstWouldAddEndPuncttrue
\mciteSetBstMidEndSepPunct{\mcitedefaultmidpunct}
{\mcitedefaultendpunct}{\mcitedefaultseppunct}\relax
\EndOfBibitem
\bibitem[Wachsmuth \latin{et~al.}(2014)Wachsmuth, Hambach, Benner, and
  Kaiser]{Wachsmuth2014}
Wachsmuth,~P.; Hambach,~R.; Benner,~G.; Kaiser,~U. {Plasmon bands in multilayer
  graphene}. \emph{Phys. Rev. B} \textbf{2014}, \emph{90}, 235434\relax
\mciteBstWouldAddEndPuncttrue
\mciteSetBstMidEndSepPunct{\mcitedefaultmidpunct}
{\mcitedefaultendpunct}{\mcitedefaultseppunct}\relax
\EndOfBibitem
\bibitem[Kramberger \latin{et~al.}(2008)Kramberger, Hambach, Giorgetti,
  Rummeli, Knupfer, Fink, Buchner, Reining, Einarsson, Maruyama, Sottile,
  Hannewald, Olevano, Marinopoulos, and Pichler]{Kramberger2008}
Kramberger,~C.; Hambach,~R.; Giorgetti,~C.; Rummeli,~M.~H.; Knupfer,~M.;
  Fink,~J.; Buchner,~B.; Reining,~L.; Einarsson,~E.; Maruyama,~S.; Sottile,~F.;
  Hannewald,~K.; Olevano,~V.; Marinopoulos,~A.~G.; Pichler,~T. Linear plasmon
  dispersion in single-wall carbon nanotubes and the collective excitation
  spectrum of graphene. \emph{Phys. Rev. Lett.} \textbf{2008}, \emph{100},
  196803\relax
\mciteBstWouldAddEndPuncttrue
\mciteSetBstMidEndSepPunct{\mcitedefaultmidpunct}
{\mcitedefaultendpunct}{\mcitedefaultseppunct}\relax
\EndOfBibitem
\bibitem[Laitenberger and Palmer(1996)Laitenberger, and
  Palmer]{Laitenberger1996}
Laitenberger,~P.; Palmer,~R.~E. Plasmon Dispersion and Damping at the Surface
  of a Semimetal. \emph{Phys. Rev. Lett.} \textbf{1996}, \emph{76},
  1952--1955\relax
\mciteBstWouldAddEndPuncttrue
\mciteSetBstMidEndSepPunct{\mcitedefaultmidpunct}
{\mcitedefaultendpunct}{\mcitedefaultseppunct}\relax
\EndOfBibitem
\bibitem[Reed \latin{et~al.}(2010)Reed, Uchoa, Joe, Gan, Casa, Fradkin, and
  Abbamonte]{Reed_2010}
Reed,~J.~P.; Uchoa,~B.; Joe,~Y.~I.; Gan,~Y.; Casa,~D.; Fradkin,~E.;
  Abbamonte,~P. The Effective Fine-Structure Constant of Freestanding Graphene
  Measured in Graphite. \emph{Science} \textbf{2010}, \emph{330},
  805--808\relax
\mciteBstWouldAddEndPuncttrue
\mciteSetBstMidEndSepPunct{\mcitedefaultmidpunct}
{\mcitedefaultendpunct}{\mcitedefaultseppunct}\relax
\EndOfBibitem
\bibitem[Gan \latin{et~al.}(2016)Gan, de~la Pe\~na, Kogar, Uchoa, Casa, Gog,
  Fradkin, and Abbamonte]{Gan_2016}
Gan,~Y.; de~la Pe\~na,~G.~A.; Kogar,~A.; Uchoa,~B.; Casa,~D.; Gog,~T.;
  Fradkin,~E.; Abbamonte,~P. Reexamination of the effective fine structure
  constant of graphene as measured in graphite. \emph{Phys. Rev. B}
  \textbf{2016}, \emph{93}, 195150\relax
\mciteBstWouldAddEndPuncttrue
\mciteSetBstMidEndSepPunct{\mcitedefaultmidpunct}
{\mcitedefaultendpunct}{\mcitedefaultseppunct}\relax
\EndOfBibitem
\bibitem[Trevisanutto \latin{et~al.}(2010)Trevisanutto, Holzmann, C\^ot\'e, and
  Olevano]{Trevisanutto_2010_b}
Trevisanutto,~P.~E.; Holzmann,~M.; C\^ot\'e,~M.; Olevano,~V. Ab initio
  high-energy excitonic effects in graphite and graphene. \emph{Phys. Rev. B}
  \textbf{2010}, \emph{81}, 121405\relax
\mciteBstWouldAddEndPuncttrue
\mciteSetBstMidEndSepPunct{\mcitedefaultmidpunct}
{\mcitedefaultendpunct}{\mcitedefaultseppunct}\relax
\EndOfBibitem
\bibitem[Yan \latin{et~al.}(2011)Yan, Thygesen, and Jacobsen]{Yan_prl2011}
Yan,~J.; Thygesen,~K.~S.; Jacobsen,~K.~W. Nonlocal Screening of Plasmons in
  Graphene by Semiconducting and Metallic Substrates: First-Principles
  Calculations. \emph{Phys. Rev. Lett.} \textbf{2011}, \emph{106}, 146803\relax
\mciteBstWouldAddEndPuncttrue
\mciteSetBstMidEndSepPunct{\mcitedefaultmidpunct}
{\mcitedefaultendpunct}{\mcitedefaultseppunct}\relax
\EndOfBibitem
\bibitem[Despoja \latin{et~al.}(2012)Despoja, Dekani\ifmmode~\acute{c}\else
  \'{c}\fi{}, \ifmmode \check{S}\else \v{S}\fi{}unji\ifmmode~\acute{c}\else
  \'{c}\fi{}, and Maru\ifmmode \check{s}\else
  \v{s}\fi{}i\ifmmode~\acute{c}\else \'{c}\fi{}]{Despoja_2012}
Despoja,~V.; Dekani\ifmmode~\acute{c}\else \'{c}\fi{},~K.; \ifmmode
  \check{S}\else \v{S}\fi{}unji\ifmmode~\acute{c}\else \'{c}\fi{},~M.;
  Maru\ifmmode \check{s}\else \v{s}\fi{}i\ifmmode~\acute{c}\else \'{c}\fi{},~L.
  Ab initio study of energy loss and wake potential in the vicinity of a
  graphene monolayer. \emph{Phys. Rev. B} \textbf{2012}, \emph{86},
  165419\relax
\mciteBstWouldAddEndPuncttrue
\mciteSetBstMidEndSepPunct{\mcitedefaultmidpunct}
{\mcitedefaultendpunct}{\mcitedefaultseppunct}\relax
\EndOfBibitem
\bibitem[Despoja \latin{et~al.}(2013)Despoja, Novko,
  Dekani\ifmmode~\acute{c}\else \'{c}\fi{}, \ifmmode \check{S}\else
  \v{S}\fi{}unji\ifmmode~\acute{c}\else \'{c}\fi{}, and Maru\ifmmode
  \check{s}\else \v{s}\fi{}i\ifmmode~\acute{c}\else \'{c}\fi{}]{Despoja_2013}
Despoja,~V.; Novko,~D.; Dekani\ifmmode~\acute{c}\else \'{c}\fi{},~K.; \ifmmode
  \check{S}\else \v{S}\fi{}unji\ifmmode~\acute{c}\else \'{c}\fi{},~M.;
  Maru\ifmmode \check{s}\else \v{s}\fi{}i\ifmmode~\acute{c}\else \'{c}\fi{},~L.
  Two-dimensional and $\ensuremath{\pi}$ plasmon spectra in pristine and doped
  graphene. \emph{Phys. Rev. B} \textbf{2013}, \emph{87}, 075447\relax
\mciteBstWouldAddEndPuncttrue
\mciteSetBstMidEndSepPunct{\mcitedefaultmidpunct}
{\mcitedefaultendpunct}{\mcitedefaultseppunct}\relax
\EndOfBibitem
\bibitem[Mowbray(2014)]{Mowbray_2014}
Mowbray,~D.~J. Theoretical electron energy loss spectroscopy of isolated
  graphene. \emph{Physica Status Solidi (b)} \textbf{2014}, \emph{251},
  2509--2514\relax
\mciteBstWouldAddEndPuncttrue
\mciteSetBstMidEndSepPunct{\mcitedefaultmidpunct}
{\mcitedefaultendpunct}{\mcitedefaultseppunct}\relax
\EndOfBibitem
\bibitem[Novko \latin{et~al.}(2015)Novko, Despoja, and \ifmmode \check{S}\else
  \v{S}\fi{}unji\ifmmode~\acute{c}\else \'{c}\fi{}]{Novko_2015}
Novko,~D.; Despoja,~V.; \ifmmode \check{S}\else
  \v{S}\fi{}unji\ifmmode~\acute{c}\else \'{c}\fi{},~M. Changing character of
  electronic transitions in graphene: From single-particle excitations to
  plasmons. \emph{Phys. Rev. B} \textbf{2015}, \emph{91}, 195407\relax
\mciteBstWouldAddEndPuncttrue
\mciteSetBstMidEndSepPunct{\mcitedefaultmidpunct}
{\mcitedefaultendpunct}{\mcitedefaultseppunct}\relax
\EndOfBibitem
\bibitem[Nazarov(2015)]{Nazarov_2015}
Nazarov,~V.~U. Electronic excitations in quasi-2D crystals: what theoretical
  quantities are relevant to experiment? \emph{New J. Phys.} \textbf{2015},
  \emph{17}, 073018\relax
\mciteBstWouldAddEndPuncttrue
\mciteSetBstMidEndSepPunct{\mcitedefaultmidpunct}
{\mcitedefaultendpunct}{\mcitedefaultseppunct}\relax
\EndOfBibitem
\bibitem[Li \latin{et~al.}(2017)Li, Ren, and He]{Li_2017}
Li,~P.; Ren,~X.; He,~L. First-principles calculations and model analysis of
  plasmon excitations in graphene and graphene/hBN heterostructure. \emph{Phys.
  Rev. B} \textbf{2017}, \emph{96}, 165417\relax
\mciteBstWouldAddEndPuncttrue
\mciteSetBstMidEndSepPunct{\mcitedefaultmidpunct}
{\mcitedefaultendpunct}{\mcitedefaultseppunct}\relax
\EndOfBibitem
\bibitem[Zheng \latin{et~al.}(2017)Zheng, Gan, Abbamonte, and
  Wagner]{Zheng_2017}
Zheng,~H.; Gan,~Y.; Abbamonte,~P.; Wagner,~L.~K. Importance of
  $\ensuremath{\sigma}$ Bonding Electrons for the Accurate Description of
  Electron Correlation in Graphene. \emph{Phys. Rev. Lett.} \textbf{2017},
  \emph{119}, 166402\relax
\mciteBstWouldAddEndPuncttrue
\mciteSetBstMidEndSepPunct{\mcitedefaultmidpunct}
{\mcitedefaultendpunct}{\mcitedefaultseppunct}\relax
\EndOfBibitem
\bibitem[Trevisanutto \latin{et~al.}(2008)Trevisanutto, Giorgetti, Reining,
  Ladisa, and Olevano]{Trevisanutto_2008}
Trevisanutto,~P.~E.; Giorgetti,~C.; Reining,~L.; Ladisa,~M.; Olevano,~V. Ab
  Initio $GW$ Many-Body Effects in Graphene. \emph{Phys. Rev. Lett.}
  \textbf{2008}, \emph{101}, 226405\relax
\mciteBstWouldAddEndPuncttrue
\mciteSetBstMidEndSepPunct{\mcitedefaultmidpunct}
{\mcitedefaultendpunct}{\mcitedefaultseppunct}\relax
\EndOfBibitem
\bibitem[Yang \latin{et~al.}(2009)Yang, Deslippe, Park, Cohen, and
  Louie]{Yang2009}
Yang,~L.; Deslippe,~J.; Park,~C.-H.; Cohen,~M.~L.; Louie,~S.~G. Excitonic
  Effects on the Optical Response of Graphene and Bilayer Graphene. \emph{Phys.
  Rev. Lett.} \textbf{2009}, \emph{103}, 186802\relax
\mciteBstWouldAddEndPuncttrue
\mciteSetBstMidEndSepPunct{\mcitedefaultmidpunct}
{\mcitedefaultendpunct}{\mcitedefaultseppunct}\relax
\EndOfBibitem
\bibitem[Yang(2011)]{Yang2011}
Yang,~L. Excitons in intrinsic and bilayer graphene. \emph{Phys. Rev. B}
  \textbf{2011}, \emph{83}, 085405\relax
\mciteBstWouldAddEndPuncttrue
\mciteSetBstMidEndSepPunct{\mcitedefaultmidpunct}
{\mcitedefaultendpunct}{\mcitedefaultseppunct}\relax
\EndOfBibitem
\bibitem[Hedin(1965)]{Hedin_1965}
Hedin,~L. New Method for Calculating the One-Particle Green's Function with
  Application to the Electron-Gas Problem. \emph{Phys. Rev.} \textbf{1965},
  \emph{139}, A796--A823\relax
\mciteBstWouldAddEndPuncttrue
\mciteSetBstMidEndSepPunct{\mcitedefaultmidpunct}
{\mcitedefaultendpunct}{\mcitedefaultseppunct}\relax
\EndOfBibitem
\bibitem[Strinati \latin{et~al.}(1982)Strinati, Mattausch, and
  Hanke]{Strinati_1982}
Strinati,~G.; Mattausch,~H.~J.; Hanke,~W. Dynamical aspects of correlation
  corrections in a covalent crystal. \emph{Phys. Rev. B} \textbf{1982},
  \emph{25}, 2867--2888\relax
\mciteBstWouldAddEndPuncttrue
\mciteSetBstMidEndSepPunct{\mcitedefaultmidpunct}
{\mcitedefaultendpunct}{\mcitedefaultseppunct}\relax
\EndOfBibitem
\bibitem[Hybertsen and Louie(1986)Hybertsen, and Louie]{Hybertsen_1986}
Hybertsen,~M.~S.; Louie,~S.~G. Electron correlation in semiconductors and
  insulators: Band gaps and quasiparticle energies. \emph{Phys. Rev. B}
  \textbf{1986}, \emph{34}, 5390--5413\relax
\mciteBstWouldAddEndPuncttrue
\mciteSetBstMidEndSepPunct{\mcitedefaultmidpunct}
{\mcitedefaultendpunct}{\mcitedefaultseppunct}\relax
\EndOfBibitem
\bibitem[Godby \latin{et~al.}(1988)Godby, Schl\"uter, and Sham]{Godby_1988}
Godby,~R.~W.; Schl\"uter,~M.; Sham,~L.~J. Self-energy operators and
  exchange-correlation potentials in semiconductors. \emph{Phys. Rev. B}
  \textbf{1988}, \emph{37}, 10159--10175\relax
\mciteBstWouldAddEndPuncttrue
\mciteSetBstMidEndSepPunct{\mcitedefaultmidpunct}
{\mcitedefaultendpunct}{\mcitedefaultseppunct}\relax
\EndOfBibitem
\bibitem[Strinati(1988)]{Strinati_1988}
Strinati,~G. Application of the Green’s functions method to the study of the
  optical properties of semiconductors. \emph{La Rivista del Nuovo Cimento
  (1978-1999)} \textbf{1988}, \emph{11}, 1--86\relax
\mciteBstWouldAddEndPuncttrue
\mciteSetBstMidEndSepPunct{\mcitedefaultmidpunct}
{\mcitedefaultendpunct}{\mcitedefaultseppunct}\relax
\EndOfBibitem
\bibitem[Onida \latin{et~al.}(2002)Onida, Reining, and Rubio]{Onida_2002}
Onida,~G.; Reining,~L.; Rubio,~A. Electronic excitations: density-functional
  versus many-body Green's-function approaches. \emph{Rev. Mod. Phys.}
  \textbf{2002}, \emph{74}, 601--659\relax
\mciteBstWouldAddEndPuncttrue
\mciteSetBstMidEndSepPunct{\mcitedefaultmidpunct}
{\mcitedefaultendpunct}{\mcitedefaultseppunct}\relax
\EndOfBibitem
\bibitem[Attaccalite and Rubio(2009)Attaccalite, and Rubio]{Attaccalite_2009}
Attaccalite,~C.; Rubio,~A. Fermi velocity renormalization in doped graphene.
  \emph{Physica Status Solidi (b)} \textbf{2009}, \emph{246}, 2523--2526\relax
\mciteBstWouldAddEndPuncttrue
\mciteSetBstMidEndSepPunct{\mcitedefaultmidpunct}
{\mcitedefaultendpunct}{\mcitedefaultseppunct}\relax
\EndOfBibitem
\bibitem[Thygesen(2017)]{Thygesen_2017}
Thygesen,~K.~S. Calculating excitons, plasmons, and quasiparticles in 2D
  materials and van der Waals heterostructures. \emph{2D Materials}
  \textbf{2017}, \emph{4}, 022004\relax
\mciteBstWouldAddEndPuncttrue
\mciteSetBstMidEndSepPunct{\mcitedefaultmidpunct}
{\mcitedefaultendpunct}{\mcitedefaultseppunct}\relax
\EndOfBibitem
\bibitem[sup()]{supp-info}
See Supplemental Material for detailed information about the experimental and
  theoretical methods and data analysis.\relax
\mciteBstWouldAddEndPunctfalse
\mciteSetBstMidEndSepPunct{\mcitedefaultmidpunct}
{}{\mcitedefaultseppunct}\relax
\EndOfBibitem
\bibitem[Kato \latin{et~al.}(2016)Kato, Minami, Koga, and Hasegawa]{Kato2016}
Kato,~R.; Minami,~S.; Koga,~Y.; Hasegawa,~M. High growth rate chemical vapor
  deposition of graphene under low pressure by RF plasma assistance.
  \emph{Carbon} \textbf{2016}, \emph{96}, 1008 -- 1013\relax
\mciteBstWouldAddEndPuncttrue
\mciteSetBstMidEndSepPunct{\mcitedefaultmidpunct}
{\mcitedefaultendpunct}{\mcitedefaultseppunct}\relax
\EndOfBibitem
\bibitem[Krivanek \latin{et~al.}(2014)Krivanek, Lovejoy, Dellby, Aoki,
  Carpenter, Rez, Soignard, Zhu, Batson, Lagos, Egerton, and
  Crozier]{Krivanek_2014}
Krivanek,~O.~L.; Lovejoy,~T.~C.; Dellby,~N.; Aoki,~T.; Carpenter,~R.~W.;
  Rez,~P.; Soignard,~E.; Zhu,~J.; Batson,~P.~E.; Lagos,~M.~J.; Egerton,~R.~F.;
  Crozier,~P.~A. Vibrational spectroscopy in the electron microscope.
  \emph{Nature} \textbf{2014}, \emph{514}, 209--212\relax
\mciteBstWouldAddEndPuncttrue
\mciteSetBstMidEndSepPunct{\mcitedefaultmidpunct}
{\mcitedefaultendpunct}{\mcitedefaultseppunct}\relax
\EndOfBibitem
\bibitem[Hachtel \latin{et~al.}(2018)Hachtel, Lupini, and Idrobo]{Hatchel_2018}
Hachtel,~J.~A.; Lupini,~A.~R.; Idrobo,~J.~C. Exploring the capabilities of
  monochromated electron energy loss spectroscopy in the infrared regime.
  \emph{Scientific reports} \textbf{2018}, \emph{8}, 5637\relax
\mciteBstWouldAddEndPuncttrue
\mciteSetBstMidEndSepPunct{\mcitedefaultmidpunct}
{\mcitedefaultendpunct}{\mcitedefaultseppunct}\relax
\EndOfBibitem
\bibitem[Giannozzi \latin{et~al.}(2020)Giannozzi, Baseggio, Bonfà, Brunato,
  Car, Carnimeo, Cavazzoni, de~Gironcoli, Delugas, Ferrari~Ruffino, Ferretti,
  Marzari, Timrov, Urru, and Baroni]{QE_2020}
Giannozzi,~P.; Baseggio,~O.; Bonfà,~P.; Brunato,~D.; Car,~R.; Carnimeo,~I.;
  Cavazzoni,~C.; de~Gironcoli,~S.; Delugas,~P.; Ferrari~Ruffino,~F.;
  Ferretti,~A.; Marzari,~N.; Timrov,~I.; Urru,~A.; Baroni,~S. Quantum ESPRESSO
  toward the exascale. \emph{J. Chem. Phys.} \textbf{2020}, \emph{152},
  154105\relax
\mciteBstWouldAddEndPuncttrue
\mciteSetBstMidEndSepPunct{\mcitedefaultmidpunct}
{\mcitedefaultendpunct}{\mcitedefaultseppunct}\relax
\EndOfBibitem
\bibitem[Perdew and Zunger(1981)Perdew, and Zunger]{LDA}
Perdew,~J.~P.; Zunger,~A. Self-interaction correction to density-functional
  approximations for many-electron systems. \emph{Phys. Rev. B} \textbf{1981},
  \emph{23}, 5048--5079\relax
\mciteBstWouldAddEndPuncttrue
\mciteSetBstMidEndSepPunct{\mcitedefaultmidpunct}
{\mcitedefaultendpunct}{\mcitedefaultseppunct}\relax
\EndOfBibitem
\bibitem[Marini \latin{et~al.}(2009)Marini, Hogan, Grüning, and
  Varsano]{yambo_2009}
Marini,~A.; Hogan,~C.; Grüning,~M.; Varsano,~D. yambo: An ab initio tool for
  excited state calculations. \emph{Comput. Phys. Commun.} \textbf{2009},
  \emph{180}, 1392 -- 1403\relax
\mciteBstWouldAddEndPuncttrue
\mciteSetBstMidEndSepPunct{\mcitedefaultmidpunct}
{\mcitedefaultendpunct}{\mcitedefaultseppunct}\relax
\EndOfBibitem
\bibitem[Sangalli \latin{et~al.}(2019)Sangalli, Ferretti, Miranda, Attaccalite,
  Marri, Cannuccia, Melo, Marsili, Paleari, Marrazzo, Prandini, Bonf{\`{a}},
  Atambo, Affinito, Palummo, Molina-S{\'{a}}nchez, Hogan, Grüning, Varsano,
  and Marini]{yambo_2019}
Sangalli,~D. \latin{et~al.}  Many-body perturbation theory calculations using
  the yambo code. \emph{J. Phys.: Condens. Matter} \textbf{2019}, \emph{31},
  325902\relax
\mciteBstWouldAddEndPuncttrue
\mciteSetBstMidEndSepPunct{\mcitedefaultmidpunct}
{\mcitedefaultendpunct}{\mcitedefaultseppunct}\relax
\EndOfBibitem
\bibitem[Ismail-Beigi(2006)]{Beigi_2006}
Ismail-Beigi,~S. Truncation of periodic image interactions for confined
  systems. \emph{Phys. Rev. B} \textbf{2006}, \emph{73}, 233103\relax
\mciteBstWouldAddEndPuncttrue
\mciteSetBstMidEndSepPunct{\mcitedefaultmidpunct}
{\mcitedefaultendpunct}{\mcitedefaultseppunct}\relax
\EndOfBibitem
\bibitem[Rozzi \latin{et~al.}(2006)Rozzi, Varsano, Marini, Gross, and
  Rubio]{Rozzi_2006}
Rozzi,~C.~A.; Varsano,~D.; Marini,~A.; Gross,~E. K.~U.; Rubio,~A. Exact Coulomb
  cutoff technique for supercell calculations. \emph{Phys. Rev. B}
  \textbf{2006}, \emph{73}, 205119\relax
\mciteBstWouldAddEndPuncttrue
\mciteSetBstMidEndSepPunct{\mcitedefaultmidpunct}
{\mcitedefaultendpunct}{\mcitedefaultseppunct}\relax
\EndOfBibitem
\bibitem[Godby and Needs(1989)Godby, and Needs]{Godby_1989}
Godby,~R.~W.; Needs,~R.~J. Metal-insulator transition in Kohn-Sham theory and
  quasiparticle theory. \emph{Phys. Rev. Lett.} \textbf{1989}, \emph{62},
  1169--1172\relax
\mciteBstWouldAddEndPuncttrue
\mciteSetBstMidEndSepPunct{\mcitedefaultmidpunct}
{\mcitedefaultendpunct}{\mcitedefaultseppunct}\relax
\EndOfBibitem
\bibitem[Venezuela \latin{et~al.}(2011)Venezuela, Lazzeri, and
  Mauri]{Venezuela2011}
Venezuela,~P.; Lazzeri,~M.; Mauri,~F. Theory of double-resonant Raman spectra
  in graphene: Intensity and line shape of defect-induced and two-phonon bands.
  \emph{Phys. Rev. B} \textbf{2011}, \emph{84}, 035433\relax
\mciteBstWouldAddEndPuncttrue
\mciteSetBstMidEndSepPunct{\mcitedefaultmidpunct}
{\mcitedefaultendpunct}{\mcitedefaultseppunct}\relax
\EndOfBibitem
\end{mcitethebibliography}

\end{document}